\documentclass[12pt]{article}
\usepackage{color}
\usepackage{graphicx}
\usepackage{rotating}
\def\LL{\left}
\def\RR{\right}
\def\beq{\begin{equation}}
\def\eeq{\end{equation}}
\def\bea{\begin{eqnarray}}
\def\eea{\end{eqnarray}}
\def\g5{\gamma_5}

\def\Re{\mathrm{Re}}
\def\Im{\mathrm{Im}}

\def\Id{\mbox{1\hspace{-1.5mm}I}}

\def\nn{\nonumber\\}

\def\ket#1{\LL|#1\RR>}

\def\nabstar#1{\nabla\kern-0.5pt\smash{\raise 4.5pt\hbox{$\ast$}}
               \kern-4.5pt_{#1}}

\def\drvstar#1{\partial\kern-0.5pt\smash{\raise 4.5pt\hbox{$\ast$}}
               \kern-5.0pt_{#1}}

\def\newline{\relax\ifhmode\null\hfil\break\else\nonhmodeerr@\newline\fi}
\def\frac#1#2{{#1\over#2}}
\def\text#1{{\hbox{\rm #1}}}
\def\flushpar{{\par \noindent}}

\def\Id{ \mbox{1\hspace{-1.2mm}I} }

\def\BE{\begin{equation}}
\def\EE{\end{equation}}
\def\BA{\begin{eqnarray}}
\def\EA{\end{eqnarray}}
\def\BAN{\begin{eqnarray*}}
\def\EAN{\end{eqnarray*}}
\def\LL{\left}
\def\RR{\right}

\def\nn{\nonumber\\}

\def\det{\mbox{det}}

\def\gm5{\gamma_5}

\input epsf.sty
\newdimen\psfigsize
\def\psfigure#1 #2 #3 #4 #5{
    \begin{figure}[tbh]
      \begin{center}
      \vbox{
        \null\vskip-0.2in\hskip#2
        \epsfxsize=#1
        \epsfbox{#4}
        \vskip -0.3in
        \caption {#5 \label{#3}}
        \vskip 0.0 true in plus 0.3 true in
      }
      \end{center}
   \end{figure}
}
\begin{document}
\thispagestyle{empty}
\begin{flushright}
NTUTH-02-505D \\
August 2002 \\
\end{flushright}
\bigskip\bigskip\bigskip
\vskip 2.5truecm
\begin{center}
{\LARGE A Linux PC cluster for lattice QCD with \\ exact chiral symmetry}
\end{center}
\vskip 1.0truecm
{
\centerline{Ting-Wai Chiu, Tung-Han Hsieh, Chao-Hsi Huang, Tsung-Ren Huang}
\vskip5mm
\centerline{Department of Physics, National Taiwan University}
\centerline{Taipei, Taiwan 106, Taiwan.}}
\centerline{\it E-mail: twchiu@phys.ntu.edu.tw}
\vskip 1cm
\bigskip \nopagebreak \begin{abstract}
\noindent

A computational system for lattice QCD with exact chiral symmetry 
is described. The platform is a home-made Linux PC cluster,
built with off-the-shelf components. At present the system constitutes 
of 64 nodes, with each node consisting of one
Pentium 4 processor (1.6/2.0/2.5 GHz), one Gbyte of PC800/1066 RDRAM,
one $ 40/80/120 $ Gbyte hard disk, and a network card. The computationally
intensive parts of our program are written in SSE2 codes. The speed of
our system is estimated to be 70 Gflops, and its price/performance ratio
is better than \$1.0/Mflops for 64-bit (double precision) computations
in quenched QCD. We discuss how to optimize its hardware and
software for computing quark propagators via the overlap Dirac operator.

\end{abstract}

\vskip 1.5cm

\newpage\setcounter{page}1


\section{Introduction}

Our objective is to extract physics from lattice QCD
with possibly minimal amount of computations.
Obviously, the required computing power exceeds
that of any desktop personal computer currently available in the market.
Thus, for one without supercomputer resources,
building a computational system \cite{Chiu:1988st}
seems to be inevitable if one really wishes to pursue a meaningful
number of any physical quantity from lattice QCD.
However, the feasibility of such a project depends not only on
the funding, but also on the theoretical advancement of the subject,
namely, the realization of exact chiral symmetry on the lattice
\cite{Kaplan:1992bt,Neuberger:1998fp}.
Now, if we also take into account of the current price/performance of
PC hardware components (CPU + RAM + hard disk\footnote{The emergence of
low-price and high-capacity ($ > 100 $ Gbyte) IDE hard disk turns out to be
also rather crucial for this project, since the data storage is enormous.}),
it seems to be the right time to rejuvenate the project \cite{Chiu:1988st}
with a new goal - to build a computational system for lattice QCD with
exact chiral symmetry.
In this paper, we outline the essential features of a Linux PC cluster
(64 nodes) which has been built at National Taiwan University.
In particular, we discuss how to optimize its hardware and software
for lattice QCD with overlap Dirac operator.

First, we start from quenched QCD calculations (i.e., ignoring any
internal quark loops by setting $ \det D = 1 $). Thus, our first
task is to compute quark propagators in the gluon field background, for
a sequence of configurations generated stochastically with
weight $ \exp( -{\cal A}_g ) $ ($ {\cal A}_g $ : pure gluon action).
Then the hardronic observables such as meson and baryon correlation functions
can be constructed, and from which the hadron masses and decay constants
can be extracted. We use the Creutz-Cabbibo-Marinari heat bath
algorithm \cite{Creutz:zw,Cabibbo:zn}
to generate ensembles of $ SU(3) $ gauge configurations.

The computation of quark propagators depends on the scheme of
lattice fermions, the hard core of lattice QCD.
In general, one requires that any quark propagator coupling to
physical hadrons must be of the form \cite{Chiu:1998eu}
\bea
 ( D_c + m_q )^{-1} \ ,
\eea
where $ m_q $ is the bare quark mass, and
$ D_c $ is a chirally symmetric and anti-hermitian
Dirac operator [ $ D_c \gamma_5 + \gamma_5 D_c = 0 $ and
$ ( i D_c )^{\dagger} = i D_c $ ].
Here we assume that $ D_c $ is doubler-free, has correct continuum
behavior, and $ D = D_c ( 1 + r a D_c )^{-1} $ is
exponentially local for smooth gauge backgrounds.
Note that the way $ m_q $ coupling to $ D_c $ is the
same as that in the continuum.
The chiral symmetry of $ D_c $ (even at finite lattice spacing)
is the crucial feature of any quark coupling to physical
hadrons. Otherwise, one could hardly reproduce the low energy strong
interaction phenomenology from lattice QCD.

For any massless lattice Dirac operator $ D $ satisfying the
Ginsparg-Wilson relation \cite{Ginsparg:1982bj}
\bea
\label{eq:gwr}
D \gamma_5 + \gamma_5 D = 2 r a D \gamma_5 D \ ,
\eea
it can be written as \cite{Chiu:1998gp}
\BAN
D = D_c ( 1 + r a D_c )^{-1} \ ,
\EAN
and the bare quark mass is naturally added to the $ D_c $ in the numerator
\cite{Chiu:1998eu},
\BAN
D(m_q) = ( D_c + m_q ) ( 1 + r a D_c )^{-1} \ .
\EAN
Then the quenched quark propagator becomes
\bea
( D_c + m_q )^{-1} = ( 1 - r m_q a )^{-1} [ D(m_q)^{-1} - r a ]
\eea

If we fix one of the end points at $ ( \vec{0}, 0 ) $
and use the Hermitcity $ D^{\dagger} = \gamma_5 D \gamma_5 $,
then only 12 (3 colors times 4 Dirac indices) columns of
\bea
\label{eq:propagator}
D(m_q)^{-1} =  D^{\dagger}(m_q) \{ D(m_q) D^{\dagger}(m_q) \}^{-1}
\eea
are needed for computing the time correlation functions of hadrons.
Now our problem is how to optimize a PC cluster
to compute $ D(m_q)^{-1} $ for a set of bare quark masses.

The outline of this paper is as follows. In Section 2, we briefly review 
our scheme of computing quark propagators via the overlap Dirac operator.
The details have been given in Ref. \cite{Chiu:2002xm}.
In Section 3, we discuss a simple scheme of memory management
for the nested conjugate gradient loops.
In Section 4, we discuss how to implement the SSE2 codes for
the computationally intense parts of our program.
In Section 5, the performance of our system is measured in terms
of a number of tests pertaining to the computation of 
quark propagators. In Section 6, we conclude with some remarks
and outlooks.


\section{Computational Scheme for quark propagators}

The massless overlap Dirac operator \cite{Neuberger:1998fp} reads as
\bea
\label{eq:overlap}
D = m_0 a^{-1} \left( \Id + \gamma_5 \frac{H_w}{\sqrt{H_w^2}} \right)
\eea
where $ H_w $ denotes the Hermitian Wilson-Dirac operator with a
negative parameter $ -m_0 $,
\bea
\label{eq:Hw}
H_w = \gamma_5 D_w = \gamma_5 (-m_0 + \gamma_\mu t_\mu + W ) \ ,
\eea
$ \gamma_\mu t_\mu $ the naive fermion operator, and $ W $
the Wilson term. Then $ D $ (\ref{eq:overlap}) satisfies the
Ginsparg-Wilson relation (\ref{eq:gwr}) with $ r = 1/(2 m_0 ) $.
In this paper, we always fix $ m_0 = 1.3 $ for our computations.
Details of our implementation 
have been given in Ref. \cite{Chiu:2002xm}.

Basically, we need to solve the following linear system
\bea
\label{eq:outer_CG}
& & D(m_q) D^{\dagger}(m_q) Y \nn
&=&\left\{ m_q^2 + \left( 2 m_0^2 - \frac{m_q^2}{2} \right)
\left[1+\frac{(\gamma_5 \pm 1)}{2} H_w \frac{1}{\sqrt{H_w^2}} \right] \right\}
Y = \Id
\eea
by conjugate gradient (CG). Then the quark propagators
can be obtained through (\ref{eq:propagator}).
With Zolotarev optimal rational approximation
\cite{Zolotarev:1877,Akhiezer:1992,vandenEshof:2001hp,Chiu:2002eh}
to $ (H_w^2)^{-1/2} $, the multiplication\footnote{Note that
the Zolotarev optimal rational polynomial in
Eq. (\ref{eq:mult_Y}) is in the form $ r^{(n,n)} $ which is
different from $ r^{(n-1,n)} $ used in Ref. \cite{Chiu:2002xm}.
We refer to Ref. \cite{Chiu:2002eh} for further discussions.}
\bea
& & H_w \left( \frac{1}{\sqrt{H_w^2}} \right) Y,
      \hspace{6mm}   h_w \equiv \frac{H_w}{\lambda_{min}}  \nn
&\simeq&  h_w ( h_w^2 + c_{2n} )
  \sum_{l=1}^{n} \frac{ b_l }{ h_w^2 + c_{2l-1} } Y
= h_w ( h_w^2 + c_{2n} )  \sum_{l=1}^{n} b_l Z_l
\label{eq:mult_Y}
\eea
can be evaluated by invoking another conjugate gradient process to
the linear systems
\bea
\label{eq:inner_CG}
( h_w^2 + c_{2l-1} ) Z_l = Y, \hspace{4mm} l = 1, \cdots, n \ .
\eea
where
\BAN
c_l &=& \frac{\mbox{sn}^2(\frac{lK'}{2n+1}; \kappa' ) }
             {1-\mbox{sn}^2(\frac{lK'}{2n+1}; \kappa' )} \\
b_l &=& d_0 \frac{ \prod_{i=1}^{n-1} ( c_{2i} - c_{2l-1} ) }
             { \prod_{i=1, i \ne l}^{n} ( c_{2i-1} - c_{2l-1} ) } \\
d_0 &=& \frac{2 \lambda }{1+ \lambda}
        \prod_{l=1}^n \frac{1+c_{2l-1}}{1+c_{2l}} \\
\lambda &=& \prod_{l=1}^{2n+1}
\frac{\Theta^2 \left(\frac{2lK'}{2n+1};\kappa' \right)}
     {\Theta^2 \left(\frac{(2l-1)K'}{2n+1};\kappa' \right)} \ .
\EAN
Here $ \Theta $ denotes the elliptic theta function,
and the Jacobian elliptic function $ \mbox{sn}( u; \kappa' ) $
is defined by the elliptic integral
\BAN
u = \int_{0}^{\mbox{sn}} \frac{dt}{\sqrt{(1-t^2)(1- \kappa'^2 t^2 )}} \ ,
\EAN
and $ K' $ is the complete elliptic integral of the first kind
with modulus $ \kappa' $,
\BAN
K' = \int_{0}^{1} \frac{dt}{\sqrt{(1-t^2)(1- \kappa'^2 t^2 )}} \ ,
\EAN
where $ \kappa' = \sqrt{ 1 - 1/b } $, $ b = \lambda_{max}^2/\lambda_{min}^2 $,
and $ \lambda_{max}^2 $ and $ \lambda_{min}^2 $ are the maximum and the
minimum of the eigenvalues of $ H_w^2 $.

Instead of solving each $ Z_l $ individually, one can use
multi-shift CG algorithm \cite{Frommer:1995ik,Jegerlehner:1996pm},
and obtain all $ Z_l $ altogether, with only a small fraction of the total
time what one had computed each $ Z_l $ separately. Evidently, one can also
apply multi-shift CG algorithm to (\ref{eq:outer_CG}) to obtain several quark
propagators with different bare quark masses.

In order to improve the accuracy of the rational approximation
as well as to reduce the number of iterations in the inner CG loop,
it is {\it crucial} to narrow the interval $ [ 1, b ] $
by projecting out the largest and some low-lying eigenmodes
of $ H_w^2 $. We use Arnoldi algorithm \cite{r:arpack}
to project these eigenmodes.
Denoting these eigenmodes by
\bea
\label{eq:eigen}
H_w u_j = \lambda_j u_j, \hspace{4mm} j = 1, \cdots, k ,
\eea
then we project the linear systems (\ref{eq:inner_CG}) to the
complement of the vector space spanned by these eigenmodes
\bea
\label{eq:inner_CG1}
( h_w^2 + c_{2l-1} ) \bar{Z_l} = \bar{Y}
\equiv ( 1 - \sum_{j=1}^k u_j u_j^{\dagger} ) Y \ ,
\hspace{4mm} l = 1, \cdots, n \ .
\eea

In the set of projected eigenvalues of $ H_w^2 $,
$ \{ \lambda_j^2 , j = 1, \cdots, k \} $,
we use $ \lambda_{max}^2 $ and $ \lambda_{min}^2 $ to denote
the least upper bound and the greatest lower bound
of the eigenvalues of $ \bar{H}_w^2 $, where
\BAN
\bar{H}_w = H_w - \sum_{j=1}^k \lambda_j u_j u_j^{\dagger} \ .
\EAN
Then the eigenvalues of
\BAN
h_w^2 = \bar{H}_w^2 / \lambda_{min}^2
\EAN
fall into the interval $ (1,b) $, $ b = \lambda_{max}^2/\lambda_{min}^2 $.

Now the matrix-vector multiplication (\ref{eq:mult_Y})
can be expressed in terms of the projected eigenmodes (\ref{eq:eigen})
plus the solution obtained from the conjugate gradient loop
(\ref{eq:inner_CG1}) in the complementary vector space, i.e.,
\bea
\label{eq:SS}
H_w \frac{1}{\sqrt{H_w^2}} Y \simeq
\frac{1}{\lambda_{min}} H_w ( h_w^2 + c_{2n} )
                        \sum_{l=1}^{n} b_l \bar{Z}_l
  + \sum_{j=1}^k  \frac{\lambda_j}{\sqrt{\lambda_j^2}}
                   u_j u_j^{\dagger} Y \equiv S
\eea
Then the breaking of exact chiral symmetry (\ref{eq:gwr}) can be measured
in terms of
\bea
\label{eq:sigma}
\sigma = \frac{ | S^{\dagger} S - Y^{\dagger} Y | }{  Y^{\dagger} Y } \ .
\eea
In practice, one has no difficulties to attain $ \sigma < 10^{-12} $
for most gauge configurations on a finite lattice \cite{Chiu:2002eh}.

Now the computation of quark propagators involves
two nested conjugate gradient loops: the so-called inner CG loop
(\ref{eq:inner_CG1}), and the outer CG loop (\ref{eq:outer_CG}).
The inner CG loop is the price what one pays for preserving the
exact chiral symmetry at finite lattice spacing.


\section{Memory management}

In this section we discuss how to configure the hardware and software
of a PC cluster such that it can attain the optimal price/performance
for the execution of the nested CG loops, (\ref{eq:outer_CG}) and
(\ref{eq:inner_CG1}).

First, we examine how much memory is required
for computing one of the 12 columns of the quark propagators
for a set of bare quark masses, since each column can be
computed independently. If the required memory can be
allocated in a single node, then each node can be assigned to work
on one of the 12 columns of the quark propagators. Then the maximum speed
of a PC cluster is attained since there is no communication overheads.
Nevertheless, the memory (RDRAM) is the most expensive component,
thus its amount should be minimized even though the maximum memory at
each node can be up to 4 Gbyte. On the other hand, if one distributes the
components of the nested CG loops across the nodes and performs parallel
computations (with MPI) through a fast network switch, then the
memory at each node can be minimal. However, the cost of a
fast network switch and its accessories is rather expensive,
and also the efficiency of the entire system will be greatly reduced
due to the communication overheads. Therefore, to optimize
the price/performance of the PC cluster relies on what is the minimal
memory required for computing one of the 12 columns of
the quark propagators.

Let $N_s = L^3 \times T$ denote the total number of lattice sites,
then each column of $ D^{-1} $ with double complex
(16 bytes) entries takes
\beq
\label{eq:Nv}
N_v = N_s \times 12 \times 16 \ \mbox{bytes}.
\eeq
Using $ N_v $ or one column as the unit,
we list the memory space of all components during
the execution of the nested CG loops :

\begin{itemize}
\item Gauge links: 3.

\item Number of projected low-lying eigenmodes: $k$

\item Quark propagators [ i.e., $ Y $ in (\ref{eq:outer_CG}) ]
      of $ N_m $ masses: $N_m/2$. \\
      (Note that each $ Y $ only takes 1/2 column since it is chiral.)

\item Conjugate gradient vectors in the CG algorithm: $N_m/2$.

\item Residual vector for the outer CG loop: $1/2$.

\item The vector $\bar{Y}_1$ (of the smallest bare quark mass) at the
      interface between the inner and the outer CG loops: 1.

\item The inner CG loop: $ 2 n + 3 $
       ( where $ n $ is the degree of Zolotarev rational polynomial), which
       consists of \\
       (i) $ \{ Z_l \} $ vectors: $ n $;  \\
       (ii) Conjugate gradient vectors $ \{ w_l \} $: $ n $; \\
       (iii) Residual vector (r): 1; \\
       (iv) $ H_w \ket{w_1}  $: 1;  \\
       (v) $ H_w^2 \ket{w_1} $: 1.
\end{itemize}

Therefore, the memory space for all components of the nested CG loops is
\bea
\label{eq:Ncg}
N_{cg} = (N_m + 1/2) + (2 n + 3) + k + 3
       = N_m + 2 n + k + 6.5 \hspace{2mm} \mbox{(columns)}
\eea
A schematic diagram of all components of the nested CG loops
is sketched in Fig. 1.

\begin{figure}[th]
\begin{center}
\includegraphics[width=14cm,height=5cm]{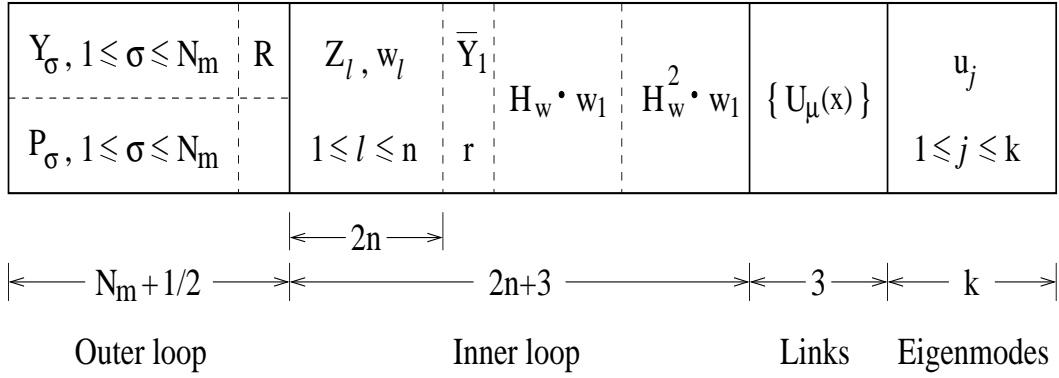}
\end{center}
\caption{
A schmatic diagram of all memory allocations for the nested CG loops.}
\end{figure}

Suppose we wish to compute quark propagators
on the $ 16^3 \times 32 $ lattice (at $ \beta = 6.0 $), with
parameters $ k = 16 $, $ n = 16 $, and $ N_m = 16 $.
Then, according to (\ref{eq:Nv}) and (\ref{eq:Ncg}),
\BAN
N_v & \simeq & 0.024 \ \mbox{Gbyte} \ , \\
N_{cg} &=&  70.5 \ \mbox{columns} \ ,
\EAN
the required memory for all components of the nested CG loops is
\BAN
N_{cg} \times N_v \simeq 70.5 \times 0.024 = 1.7  \ \mbox{Gbyte}
\EAN

This seems to imply that one should
install\footnote{At present, most Pentium 4 motherboards designed for
housing PC800 RDRAM have 4 memory slots.}
four stripes of $ 512 $ Mbyte modules (i.e. total 2 Gbyte) at each node,
if one wishes to let each node compute independently, and to
attain the maximum speed of the PC cluster.
However, this is a rather expensive solution at this moment,
in view of the current price of $ 512 $ Mbyte modules.
On the other hand, if one distributes the components of the nested CG
loops across the nodes and performs parallel computations (with MPI)
through a fast network switch, then the price/performance seems
to be even worse than the former solution.

Fortunately, we observe that {\it not} all column vectors are used
simultaneously at any step of the nested CG loops, and also the
computationally intense part is at the inner CG loop.
Thus we can use the hard disk as the virtual memory for the storage of
the intermediate solution vectors and their conjugate gradient vectors
($ Y_{\sigma}, P_{\sigma}, \sigma = 1, ... N_{m} $)
at each iteration of the outer CG loop, while the CPU is
working on the inner CG loop. Then the minimal physical memory required
at each node can be greatly reduced.
Also, the projected eigenmodes are not required to be
kept inside the memory, since they are only needed
at the start of the inner CG loop to compute
$ \bar{Y}_1 $ (for the smallest bare quark mass),
\BAN
\bar{Y}_1 \equiv ( 1 - \sum_{j=1}^k u_j u_j^{\dagger} ) Y_1 \ ,
\EAN
and
\BAN
\sum_{j=1}^k \frac{\lambda_j}{\sqrt{\lambda_j^2}} u_j u_j^{\dagger} Y_1
\equiv {\varepsilon}_p Y_1
\EAN
where $ {\varepsilon}_p Y_1 $ is only needed for computing $ S $
(\ref{eq:SS}) at the completion of the inner CG loop. Thus one has the
options to keep the vector $ {\varepsilon}_p Y_1 $ inside the memory during
the entire inner CG loop or save it to the hard disk and then retrieve it
at the completion of the inner CG loop. Further, since $ \bar{Y}_1 $ is
only needed at the start of the inner CG loop, so it can share the same
memory location with the residual vector $ r $.

Now it is clear that the minimum memory at each node
(without suffering a substantial loss in the performance) is
\BAN
\label{eq:Ncg_min}
N_{cg}^{min} = ( 2n + 3 ) + 3 = 2n + 6 \ \mbox{(columns)} \ ,
\EAN
which suffices to accommodate the link variables and all
relevant vectors for the inner CG loop.
After the completion of the inner CG loop and the vector $ S $ (\ref{eq:SS})
is computed, the memory space of $ 2n + 3 $ column vectors is
released, and the vectors $ \{ Y_{\sigma} \} $ and $ \{ P_{\sigma} \} $
of the outer CG loop can be read from the hard disk, which are then
updated to new values according to the CG algorithm.

With this simple scheme of memory management, the minimal memory
for computing one of the 12 columns of the quark propagators
(for a set of bare quark masses) on the $ 16^3 \times 32 $ lattice
with $ n=16 $ (degree of Zolotarev rational polynomial) becomes
\BAN
N_{cg}^{min} \times N_v = 38 \times 0.024 = 0.912 \ \mbox{Gbyte}.
\EAN
Thus the computation can be performed at a single node with one Gbyte
of memory, which can be implemented by installing four stripes of 256
Mbyte memory modules, a much more economic solution than using
$ 4 \times 512 $ Mbyte modules. Moreover, the time for disk I/O
(at the interface of inner and outer CG loops)
only constitutes a few percent of the total time for the execution
of the entire nested CG loops (Table \ref{tab:diskIO}).
This is the optimal memory configuration for a PC cluster to
compute quark propagators on the $ 16^3 \times 32 $ lattice,
which of course is not necessarily the optimal one for other lattice sizes.
However, our simple scheme of memory management for the nested
CG loops should be applicable to any lattice sizes, as well as to
other systems.

In passing, we emphasize that the Zolotarev optimal rational approximation
to $ (H_w^2)^{-1/2} $ plays a crucial role to minimize the
number of vectors required for the inner CG loop. If one had used
other rational approximations, then it would require a very large $ n $
to preserve exact chiral symmetry to a high precision
(e.g., $ \sigma < 10^{-11} $). In that case, it would be impossible
to attain the optimal price/performance as what has been outlined above.



\section{The SSE2 acceleration}

With the optimal memory allocation for each node, we further
enhance the performance of our lattice QCD codes (in Fortran)
by rewriting its computationally intense parts in the SSE2 assembly
codes of Pentium 4. In this section, we briefly review the basic
features of the vector unit (SSE2) of Pentium 4, and then describe
how to implement SSE2 codes in our lattice QCD program.

\subsection{The basic features of SSE2}

\begin{figure}
\centerline{\includegraphics[width=8cm,height=4cm]{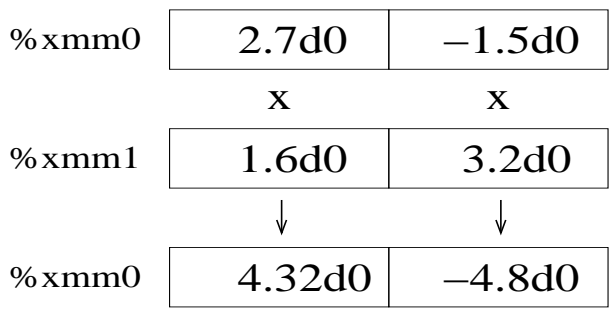}}
\caption{Double precision multiplication performed by the
         SSE2 instruction in the SIMD registers.}
\label{f:mulpd}
\end{figure}

The simplest and the most efficient scheme of parallel computation
is Single Instruction Multiple Data (SIMD).
It can be implemented inside CPU through a set of long registers.
If each register can accommodate several (say, $ s $) data entries,
then any operation (addition, subtraction, multiplication and division)
on these registers will act on all data entries in parallel,
thus yields the speed-up by a factor of $ s $ comparing with normal registers.
A schematic diagram is shown in Fig. \ref{f:mulpd}.

Even though Intel had implemented the vector unit in their
CPUs since Pentium-MMX series, only in the most recent IA-32
Pentium 4 and the advanced IA-64 Itanium, the architecture
has been extended to SSE2 (Streamed SIMD Extension 2) to
incorporate double precision data entries.

The Pentium 4 processor has eight registers
( {\textbf \%xmm0, \%xmm1, \ldots, \%xmm7} ) for SIMD operations
\cite{r:intel:sse2}.
Each register is 128 bits wide and can accomodate 4 integers, or
4 single-precision or 2 double-precision floating point numbers.
Since we always use double precision floating point numbers in our
program, the execution speed of our program can be almost
doubled if SSE2 is turned on judiciously in the computationally
intensive parts.
Note that SSE2 complies with the IEEE 32-bit and 64-bit
arithmetic, thus the precision is lower than the extended 80-bit
precision of the normal registers in Pentium 4.
However, the difference is less than one part in $ 10^{15} $
(double precision), thus is negligible in our computations.

\subsection{How to implement SSE2 codes in Fortran programs}

Since our lattice QCD codes were originally written in Fortran 77,
it would be natural if SSE2 codes can be directly embedded in
our Fortran program.
However, to our knowledge, the Fortran compilers currently available
in the market do not support the option of inlining SSE2 codes.
Moreover, for optimal performance of SSE2, the data should be aligned to
16-byte memory boundary. This can be easily carried out in C.
Therefore our strategy to implement SSE2 codes is rewrite the main
program unit in C such that the data arrays are allocated and aligned to
16 bytes memory boundary, then the SSE2 codes are embedded in C
subroutines which are then called by original routines in Fortran.

Of course, if one has written lattice QCD codes in C, then
the SSE2 codes can be embedded in C routines directly, without
dealing with the interface of C and Fortran.

In the following, we illustrate our scheme of implementing SSE2 codes
with an example program. The default compilers are $ {\bf gcc} $
and $ {\bf g77} $ in Linux.

\begin{verbatim}
        program main
        implicit none
        integer n
        parameter (n=100)
        double precision r(n), v(n)
        call vxzero(n, r, v)
        end

        subroutine vxzero(n, r, v)
        implicit none
        integer n
        double precision c, r(*), v(*)
        ...
        call vadd(n, c, r, v)             !  r = r + c v
        ...                               !  c:    scalar
        end                               !  r, v: vector

\end{verbatim}
Here the Fortran {\bf main} program calls the subroutine {\bf vxzero} which
in turn calls a computationally intensive routine {\bf vadd}.

First, we rewrite the main program in C, with the data arrays allocated
and properly aligned.

\begin{verbatim}
        #include <malloc.h>
        int main(int argc, char **argv)
        {
            int n=100;
            double *r, *v;
        /* setup the environment for Fortran */
            f_setarg(argc, argv);
            f_setsig();
            f_init();
        /* allocate r & v, and align them to 16-byte boundary */
            r = memalign(16, n*sizeof(double));
            v = memalign(16, n*sizeof(double));
        /* call the Fortran subroutine */
            vxzero_(&n, r, v);
        /* shutdown the I/O channels of Fortran */
            f_exit();
            exit(0);
            return 0;
        }
\end{verbatim}
The function call {\bf memalign()} dynamically allocates 16 bytes
aligned pointers {\bf r} and {\bf v}.
Then the aligned arrays {\bf v[\,]} and {\bf r[\,]} can be passed
to C subroutines for SSE2 operations.

Next we rewrite the computationally intensive routine {\bf vadd}
in C with embedded SSE2 codes.

\begin{verbatim}
        /* load variable a into %%xmm0 */
        #define sse_load(a)  \
            __asm__ __volatile__ ( "movapd %0, %%xmm0" :: "m" (a))

        /* r = r + %%xmm0 x v          */
        #define sse_add(r, v)               \
            __asm__ __volatile__ (          \
                "movapd %1, %%xmm1 \n\t"    \
                "movapd %2, %%xmm2 \n\t"    \
                "mulpd %%xmm0, %%xmm2 \n\t" \
                "addpd %%xmm1, %%xmm2 \n\t" \
                "movapd %%xmm2, %0"         \
                :                           \
            /* store to address (r), which is indexed as %0 */  \
                "=m" (r)                    \
                :                           \
            /* load from address (r) and (v), which are  \
               indexed as %1 and %2, respectively */     \
                "m" (r), "m" (v))

        #define ALIGN16 __attribute__ ((aligned (16)))

        void vadd_(int *n, double *coeff, double *r, double *v)
        {
            int i, len;
            static double cc[2] ALIGN16;

          /* the array cc is aligned to 16-byte boundary */

            cc[0] = cc[1] = *coeff;
            sse_load(cc[0]);
            len = (*n)/2;
            for (i=0; i<len*2; i+=2) {
                sse_add(r[i], v[i]);
            }
            if (*n % 2 != 0)
                r[len*2] = r[len*2] + cc[0] * v[len*2];
        }
\end{verbatim}

Note that we have added the keyword "{\_\_volatile\_\_}"
( an GNU extension ) in the macro "{\_\_asm\_\_}".
Its purpose is to ensure that the compiler does not rearrange
the order of execution of the codes during compilation.
Finally, all object modules are linked by $ {\bf gcc} $ with the option
"-lg2c".

\subsection{The implementation of $ H_w $ times $ \ket{v}$}

In our lattice QCD program, most of the execution time is spent in
solving quark propagators via the nested CG loops.
Thus the execution time is dominated by the operation
$ H_w $ times  $ \LL|v\RR> $, which
is performed many times ( $ > 10^5 $ in most cases ) before
the final results of quark propagators can be obtained.
Thus it is crucial to optimize this operation with SSE2 codes.

First, we have to set up the correspondence between the data structures
used by C and Fortran routines in our program, in particular, for the
link variables and the relevant vectors in the nested CG loops.

Suppose we write the arrays of link variables and a
column vector $ v $ in the syntax of Fortran as
\BAN
u(i,j,\mu,x)\,, \\
v(i,k,x)\,,
\EAN
where $i$ and $j$ are the color indices,
$\mu$ is the space-time direction,
$k$ is the spinor index, and $x$ is the site index.
Now the question is how to access the elements of these arrays
in C routines. To resolve this problem, we define some data structures
in C as follows.
\begin{verbatim}
    /* SU(3) matrix, (c01,c02) forms the complex number of u11, and
                     (c03,c04) of u21, etc.                           */
    typedef struct {
        double c01, c02, c03, c04, c05, c06;
        double c07, c08, c09, c10, c11, c12;
        double c13, c14, c15, c16, c17, c18;
    } su3_t;

    /* there are 4 link variables at each site. */
    typedef struct {
        su3_t mu1, mu2, mu3, mu4;
    } ulink_t;

    /* SU(3) vector, (c1,c2) forms the complex number of v1,
                     (c3,c4) of v2, and (c5,c6) of v3. */
    typedef struct {
        double c1, c2, c3, c4, c5, c6;
    } vector_t;

    /* SU(3) Dirac spinor. */
    typedef struct {
        vector_t s1, s2, s3, s4;
    } spinor_t;
\end{verbatim}

Then the correspondence can be easily established.
For example, the elements $u(3,2,1,x)$ and
$v(2,4,x)$ can be accessed by C routines as
($u[x]$.mu1.c11, $u[x]$.mu1.c12) and
( $v[x]$.s4.c3, $v[x]$.s4.c4 ) respectively.

Now we rewrite $ H_w $ as
\BAN
H_w(x,y) = \gamma_5 \left\{ (4-m_0) \delta_{x,y} +
\frac{1}{2} \sum_{\mu=1}^4
[ (-1+\gamma_\mu) U_\mu(x) \delta_{x+\mu,y} -
  (1+\gamma_\mu) U^{\dagger}_\mu (x-\mu) \delta_{x-\mu,y} ] \right\}
\EAN

Then the multiplication of $ H_w $ to a column vector $ \LL|v\RR> $
can be optimized by minimizing the number of multiplications
involving the link variables.
For example, the multiplication in $ ( -1 + \gamma_1 ) u \LL|v\RR> $
can be written as ( in the spinor space )
\BAN
(-\Id + \gamma_1) u \LL|v\RR> =
	\LL(\begin{array}{c} r_1 \\ r_2 \\ r_3 \\ r_4 \end{array}\RR) =
	\LL(\begin{array}{cccc}
	-u & 0 & 0 & u \\
	0 & -u & u & 0 \\
	0 & u & -u & 0 \\
        u & 0 & 0 & -u \end{array}\RR)
	\LL(\begin{array}{c} v_1 \\ v_2 \\ v_3 \\ v_4 \end{array}\RR) =
        \LL(\begin{array}{c} u(v_4-v_1) \\ u(v_3-v_2) \\ -r_2 \\ -r_1
		\end{array}\RR)
\EAN
where all indices are suppressed except the spinor indices.
It is clear that the vectors $ v_4-v_1 $ and $ v_3-v_2 $ should be computed
first, before they are multiplied by link variable $ u $ ( generic symbol
for $ U_\mu/2 $ ).
For example, the operation $ v_4-v_1$ can be performed by the following
macros with SSE2.
\begin{verbatim}
#define mvpv(v1, v2) \
    __asm__ __volatile__ ( \
        "movapd %0, %%xmm0 \n\t" \
        "movapd %1, %%xmm1 \n\t" \
        "movapd %2, %%xmm2 \n\t" \
        "subpd %3, %%xmm0 \n\t" \
        "subpd %4, %%xmm1 \n\t" \
        "subpd %5, %%xmm2" \
        : : \
        "m" ((v2).c1), \
        "m" ((v2).c3), \
        "m" ((v2).c5), \
        "m" ((v1).c1), \
        "m" ((v1).c3), \
        "m" ((v1).c5))
\end{verbatim}
Similarly, we have
\BAN
(-\Id - \gamma_1) u^{\dagger} \LL|v\RR> &=&
	\LL(\begin{array}{c} r_1 \\ r_2 \\ r_3 \\ r_4 \end{array}\RR) =
        \LL(\begin{array}{c} -u^{\dagger}(v_4+v_1) \\ -u^{\dagger}(v_3+v_2) \\ r_2 \\ r_1
		\end{array}\RR)\,, \\
(-\Id + \gamma_2) u \LL|v\RR> &=&
	\LL(\begin{array}{c} r_1 \\ r_2 \\ r_3 \\ r_4 \end{array}\RR) =
	\LL(\begin{array}{c} -u(v_1+iv_4) \\ -ir_3 \\ -u(v_3+iv_2) \\ -ir_1
		\end{array}\RR)\,, \\
(-\Id - \gamma_2) u^{\dagger} \LL|v\RR> &=&
	\LL(\begin{array}{c} r_1 \\ r_2 \\ r_3 \\ r_4 \end{array}\RR) =
        \LL(\begin{array}{c} -ir_4 \\ -u^{\dagger}(v_2+iv_3) \\ -ir_2 \\ -u^{\dagger}(v_4+iv_1)
		\end{array}\RR)\,, \\
(-\Id + \gamma_3) u \LL|v\RR> &=&
	\LL(\begin{array}{c} r_1 \\ r_2 \\ r_3 \\ r_4 \end{array}\RR) =
	\LL(\begin{array}{c} u(v_3-v_1) \\ -u(v_2+v_4) \\ -r_1 \\ r_2
		\end{array}\RR)\,, \\
(-\Id - \gamma_3) u^{\dagger} \LL|v\RR> &=&
	\LL(\begin{array}{c} r_1 \\ r_2 \\ r_3 \\ r_4 \end{array}\RR) =
        \LL(\begin{array}{c} -u^{\dagger}(v_1+v_3) \\ u^{\dagger}(v_4-v_2) \\ r_1 \\ -r_2
		\end{array}\RR)\,, \\
(-\Id + \gamma_4) u \LL|v\RR> &=&
	\LL(\begin{array}{c} r_1 \\ r_2 \\ r_3 \\ r_4 \end{array}\RR) =
	\LL(\begin{array}{c} -ir_3 \\ -ir_4 \\ -u(v_3+iv_1) \\ -u(v_4+iv_2)
		\end{array}\RR)\,, \\
(-\Id - \gamma_4) u^{\dagger} \LL|v\RR> &=&
	\LL(\begin{array}{c} r_1 \\ r_2 \\ r_3 \\ r_4 \end{array}\RR) =
        \LL(\begin{array}{c} -u^{\dagger}(v_1+iv_3) \\ -u^{\dagger}(v_2+iv_4) \\ -ir_1 \\ -ir_2
		\end{array}\RR)\,.
\EAN
So the multiplications involving the link variables can be implemented as
\begin{verbatim}
    /* for each lattice size */
    for (x=0; x<ldim; x++) {
    /* prefetch for the current multiplication */
        y = iup[x].mu1-1;
        _prefetch_su3(&(u[x].mu1));
        _prefetch_spinor(&(v[y]));
    /* prefetch for the next multiplication */
        z = idn[x].mu1-1;
        _prefetch_su3(&(u[z].mu1));
        _prefetch_spinor(&(v[z]));
    /* r1.s1 = u[x].mu1 * (v[y].s4 - v[y].s1) */
        mvpv(v[y].s1, v[y].s4);
        su3mul(r1.s1, u[x].mu1);
    /* r1.s2 = u[x].mu1 * (v[y].s3 - v[y].s2) */
        mvpv(v[y].s2, v[y].s3);
        su3mul(r1.s2, u[x].mu1);
    /* r1.s3 = -r1.s2 */
        mvset(r1.s3, r1.s2);
    /* r1.s4 = -r1.s1 */
        mvset(r1.s4, r1.s1);

    /* prefetch for the next multiplication */
        y = iup[x].mu2-1;
        _prefetch_su3(&(u[x].mu2));
        _prefetch_spinor(&(v[y]));
    /* r2.s1 = -(u[x].mu1)^{\dagger} * (v[y].s1 + v[y].s4) */
        mvmv(v[z].s1, v[z].s4);
        su3Hmul(r2.s1, u[z].mu1);
    /* r2.s2 = -(u[x].mu1)^{\dagger} * (v[y].s2 + v[y].s3) */
        mvmv(v[z].s2, v[z].s3);
        su3Hmul(r2.s2, u[z].mu1);
    /* r2.s3 = r2.s2 */
        pvset(r2.s3, r2.s2);
    /* r2.s4 = r2.s1 */
        pvset(r2.s4, r2.s1);
        ...
\end{verbatim}
where {\bf r1}, {\bf r2}, \ldots, and {\bf v[\,]} are declared as the type
{\bf spinor\_t}, and {\bf u[\,]} is declared as the type {\bf ulink\_t}.
Note that prefetching has been inserted in order to attain the
optimal performance.
Finally, we have 8 vector segments {\bf r1}, \ldots, {\bf r8}, and a diagonal
term. They are summed over to give the final result of $v[y]$,
\BAN
v[y].s1 &=& r1.s1 + r2.s1 + r3.s1 + r4.s1 + r5.s1 + r6.s1 + r7.s1 + r8.s1
	\nonumber \\
        && + (4-m_0) * v[x].s1\,, \nonumber\\
v[y].s2 &=& r1.s2 + r2.s2 + r3.s2 + r4.s2 + r5.s2 + r6.s2 + r7.s2 + r8.s2
	\nonumber \\
        && + (4-m_0) * v[x].s2\,, \nonumber\\
v[y].s3 &=& -(r1.s3 + r2.s3 + r3.s3 + r4.s3 + r5.s3 + r6.s3 + r7.s3 + r8.s3
	\nonumber \\
        && + (4-m_0) * v[x].s3)\,, \nonumber\\
v[y].s4 &=& -(r1.s4 + r2.s4 + r3.s4 + r4.s4 + r5.s4 + r6.s4 + r7.s4 + r8.s4
	\nonumber \\
        && + (4-m_0) * v[x].s4)\,, \nonumber
\EAN

Next we come to the question how to implement SSE2 codes for a
$SU(3)$ matrix times a vector, the most crucial part in
$ H_w $ times $ \LL|v\RR>$.
This problem has been solved by L\"uscher \cite{Luscher:2001tx},
and his SSE2 codes is available in the public domain \cite{FermiQCDweb}.
We found that L\"uscher's code is quite efficient, and
have adopted it in our program.
For completeness, we briefly outline L\"uscher's algorithm as follows.

Consider
\BAN
\LL( \begin{array}{ccc}
        u_{11} & u_{12} & u_{13} \\
        u_{21} & u_{22} & u_{23} \\ 
        u_{31} & u_{32} & u_{33} \end{array} \RR) \times
\LL( \begin{array}{c}
        y_1 \\ y_2 \\ y_3 \end{array} \RR) =
\LL( \begin{array}{c}
        r_1 \\ r_2 \\ r_3 \end{array} \RR)\\.
\EAN

First, the elements ($ y_1, y_2, y_3 $) of the vector $ \LL|y\RR> $
are copied to the registers \%xmm0, \%xmm1, and \%xmm2, respectively.
Then the real part of the SU(3) matrix $ \{ u_{mn} \} $ is read
sequentially, and is multiplied to $ \LL|y\RR> $
at \%xmm0, \%xmm1, and \%xmm2, and the result is stored
at \%xmm3, \%xmm4, and \%xmm5,
\BAN
\%\mathrm{xmm0} &=& (\Re(y_1), \Im(y_1))\,, \nn
\%\mathrm{xmm1} &=& (\Re(y_2), \Im(y_2))\,, \nn
\%\mathrm{xmm2} &=& (\Re(y_3), \Im(y_3))\,, \nn
\%\mathrm{xmm3} &=& (t_1, t_2)\,, \nn
\%\mathrm{xmm4} &=& (t_3, t_4)\,, \nn
\%\mathrm{xmm5} &=& (t_5, t_6)\,, \nonumber
\EAN
where
\bea
t_1 &=&\Re(u_{11})\Re(y_1)+\Re(u_{12})\Re(y_2)+\Re(u_{13})\Re(y_3)\,,\nn
t_2 &=&\Re(u_{11})\Im(y_1)+\Re(u_{12})\Im(y_2)+\Re(u_{13})\Im(y_3)\,,\nn
t_3 &=&\Re(u_{21})\Re(y_1)+\Re(u_{22})\Re(y_2)+\Re(u_{23})\Re(y_3)\,,\nn
t_4 &=&\Re(u_{21})\Im(y_1)+\Re(u_{22})\Im(y_2)+\Re(u_{23})\Im(y_3)\,,\nn
t_5 &=&\Re(u_{31})\Re(y_1)+\Re(u_{32})\Re(y_2)+\Re(u_{33})\Re(y_3)\,,\nn
t_6 &=&\Re(u_{31})\Im(y_1)+\Re(u_{32})\Im(y_2)+\Re(u_{33})\Im(y_3)\,.\nonumber
\eea

Next, multiply the vector $y$ by $i=(0,1)$, i.e.,
\BAN
\begin{array}{lclcl}
\%\mathrm{xmm0} &\rightarrow& (\Im(y_1), \Re(y_1))
                &\rightarrow& (-\Im(y_1), \Re(y_1))\,, \\
\%\mathrm{xmm1} &\rightarrow& (\Im(y_2), \Re(y_2))
                &\rightarrow& (-\Im(y_2), \Re(y_2))\,, \\
\%\mathrm{xmm2} &\rightarrow& (\Im(y_3), \Re(y_3))
                &\rightarrow& (-\Im(y_3), \Re(y_3))\,,
\end{array}
\EAN
which is implemented by the following SSE2 code
\begin{verbatim}
    static int sn3[4] ALIGN16 = {0x0,0x80000000,0x0,0x0};
    #define su3mul(r, u) \
        ...                     \
        "xorpd %9, %%xmm0 \n\t" \
        "xorpd %9, %%xmm1 \n\t" \
        "xorpd %9, %%xmm2 \n\t" \
        ...                     \
        ::                      \
        ...                     \
        "m" (sn3[0]));
\end{verbatim}
Then the imaginary part of $ \{ u_{mn} \} $ is read and
multiplied to $ i y $, and the final result is
\BAN
\%\mathrm{xmm3} &=& (t_1+s_1, t_2+s_2)\,, \\
\%\mathrm{xmm4} &=& (t_3+s_3, t_4+s_4)\,, \\
\%\mathrm{xmm5} &=& (t_5+s_5, t_6+s_6)\,,
\EAN
where
\bea
s_1 &=& -\Im(u_{11})\Im(y_1)-\Im(u_{12})\Im(y_2)-\Im(u_{13})\Im(y_3)\,, \nn
s_2 &=& +\Im(u_{11})\Re(y_1)+\Im(u_{12})\Re(y_2)+\Im(u_{13})\Re(y_3)\,, \nn
s_3 &=& -\Im(u_{21})\Im(y_1)-\Im(u_{22})\Im(y_2)-\Im(u_{23})\Im(y_3)\,, \nn
s_4 &=& +\Im(u_{21})\Re(y_1)+\Im(u_{22})\Re(y_2)+\Im(u_{23})\Re(y_3)\,, \nn
s_5 &=& -\Im(u_{31})\Im(y_1)-\Im(u_{32})\Im(y_2)-\Im(u_{33})\Im(y_3)\,, \nn
s_6 &=& +\Im(u_{31})\Re(y_1)+\Im(u_{32})\Re(y_2)+\Im(u_{33})\Re(y_3)\,. \nonumber
\eea

\begin{table}
\begin{center}
\begin{tabular}{c|c|c|c}
Lattice\ Size &  SSE2 off &  SSE2 on & speed-up \\
\hline
\hline
$8^3 \times 24$  & 0.034   & 0.018  & 1.89  \\
$10^3\times 24$  & 0.065   & 0.036  & 1.81  \\
$12^3\times 24$  & 0.110   & 0.063  & 1.75  \\
$16^3\times 32$  & 0.328   & 0.183  & 1.79  \\
\end{tabular}
\caption{
The execution time (in unit of second) for $ H_w $ multiplying
a column vector $ Y $, with SSE2 turned on and off.
The test is performed at a Pentium 4 (2 GHz) node.
}
\label{tab:SSE2}
\end{center}
\end{table}


\section{Performance of the system}

In this section, we measure the performance of our system
by a number of tests pertaining to the computation of
quark propagators.

In Table \ref{tab:SSE2}, we list the execution time (in unit of second)
for $ H_w $ multiplying a column vector $ Y $, for both cases with SSE2
turned on and off, and for several lattice sizes.
The data shows that turning on SSE2 can speed up our program by
a factor $ \sim 1.8 $.

In Table \ref{tab:ncv}, we list the execution time (in unit of second)
for projecting 20 low-lying eigenmodes of $ H_w^2 $ using ARPACK,
versus the number of Arnoldi vectors. It is clear that there exists
an optimal number of Arnoldi vectors for a projection,
which of course depends on the gauge configuration.
In Table \ref{tab:ncv}, the optimal number is $ \sim 100 $,
which amounts to $ \sim 240 $ Mbyte for the $ 8^3 \times 24 $ lattice.
However, for larger lattices such as $ 16^3 \times 32 $, the
optimal number may require more than
one gigabyte of memory. In this case, the projection
of eigenmodes is carried out at some nodes with 2 gigabyte of memory.

\begin{table}
\begin{center}
\begin{tabular}{c|c|c}
Arnoldi vectors &  Iterations &  Time  \\
\hline
\hline
40  &    756      &      12923   \\
50  &    160      &       4757   \\
60  &    108      &       4730   \\
70  &     82      &       4414   \\
80  &     65      &       4131   \\
90  &     55      &       4103   \\
100 &     46      &       4100   \\
120 &     37      &       4251   \\
140 &     32      &       4869   \\
160 &     26      &       5002
\end{tabular}
\caption{The execution time (in unit of second) for projecting $ 20 $
         low-lying eigenmodes of $ H_w^2 $ using ARPACK, versus the number
         of Arnoldi vectors. The test is performed at a
         Pentium 4 (1.6 GHz) node, for a gauge configuration on the
         $8^3 \times 24$ lattice, at $\beta=5.8$.
         Each eigenmode satisfies
         $ \| (H_w^2-\lambda^2) |x \rangle \| < 10^{-13}$.}
\label{tab:ncv}
\end{center}
\end{table}


In Table \ref{tab:diskIO}, we measure the time
used by disk I/O in our simple scheme of memory management for
the nested CG loops, versus the number of bare quark masses.
The test is performed at a Pentium 4 (2 GHz) node,
for the $16^3 \times 32$ lattice, and with 16 projected eigenmodes.
The disk I/O time is the difference of the total execution
time between two cases of turning on and off of the memory management.
It is remarkable that the percentage of disk I/O time is only
$ 3\% $ of the total execution time even for 16 bare quark masses, and
with 16 projected eigenmodes. Evidently, for the $ 16^3 \times 32 $ lattice,
our simple scheme of memory management is more efficient
and less expensive than any other options, e.g.,
parallel computing (with MPI) through a fast network switch.

\begin{table}
\begin{center}
\begin{tabular}{c|ccc}
$N_m$       & 1 & 8 & 16 \\
\hline
\hline
CG time         & 491.9  &   494.8 &  497.1   \\
disk I/O time   & 7.6    &   8.9   &  14.3    \\
Total time      & 499.5  &   503.7 &  511.4   \\
disk I/O (\%)   & 1.5\%  &   1.8\% &  2.9\%
\end{tabular}
\caption{The percentage of time spent in memory management
(disk I/O) versus the number of bare quark masses ($N_m$).
The test is performed at a Pentium 4 (2 GHz) node,
for the $16^3 \times 32$ lattice, and with 16 projected eigenmodes.
The time (in unit of second) shown here is only for completing
one outer CG iteration for one column of $ D^{-1} $.
}
\label{tab:diskIO}
\end{center}
\end{table}

In Table \ref{tab:perf1}, we list the execution time (in unit of second)
for a Pentium 4 (2 GHz) node to compute 12 columns of 
quark propagators in a topologically nontrivial gauge background
at $ \beta = 5.8 $ on the $ 8^3 \times 24 $ lattice,
versus the number of projected low-lying eigenmodes.
Other parameters for the test are:
the degree of Zolotarev rational polynomial is $ n = 16 $;
the number of bare quark masses is $ N_m = 12 $;
each projected eigenmode satisfies
             $ \|(H_w^2-\lambda^2) |x\rangle \|< 10^{-13}$,
and the stopping criterion for inner and outer CG loops is
$ \epsilon = 10^{-11} $.
The execution time is decomposed into three parts :
(i) the projections of high and low-lying eigenmodes\footnote{Note that
the projection time listed in the 2nd column of Table \ref{tab:perf1}
includes 167 seconds for projecting 4 highest eigenmodes
of $ H_w^2 $.}; (ii) computing 12 columns of
$ ( D D^{\dagger} )^{-1} $ via the nested CG loops;
and (iii) computing $ D^{\dagger} $
and multiplying it to $ ( D D^{\dagger} )^{-1} $. The total time
is listed in the last column of the table.
For completeness, we also list $ \lambda_{max} $ and $ \lambda_{min} $
of $ | \bar{H_w} | $ (after the projections), the total numbers of iterations
of the outer CG loop and average iterations of the inner CG loop,
as well as the precision of exact chiral symmetry in terms of
$ \sigma $ (\ref{eq:sigma}).
Evidently, the time for projecting out the high and low-lying
eigenmodes is only a very small fraction of the total execution time
for computing 12 columns of quark propagators.
However, the projections have very
significant impacts on the total execution time since it
yields the speed-up by a factor of 2.44, as comparing the
first row (no projections) with the last row
(projections of 40 low-lying eigenmodes). Moreover, with projections,
the exact chiral symmetry can be easily preserved to a very high
precision ($ \sigma < 10^{-13} $).
This suggests that {\it one should project as many low-lying eigenmodes
as possible}, before executing the nested CG loops. In general, we suspect
that the optimal number of projections depends on the projection
algorithm, the amount of memory of the system, as well as
the gauge configuration.

\begin{table}
\begin{center}
\begin{tabular}{cc|cccccccc}
\multicolumn{2}{c}{projections} & & & inner CG & outer CG
      & $\chi$ sym. &
        \multicolumn{1}{c}{CG} & $ D^{\dagger} $ mult. & Total\\
\# & time & $\lambda_{min} $ & $ \lambda_{max}$
      &ave. iters.&tot. iters.& $\sigma$(max.) & time &time& time \\
\hline
\hline
 0 & 0     & 0.017 & 6.207 & 965 & 1282
           & $ 4.5 \times 10^{-10}$ & 137221 & 15070 & 152291 \\
 8 & 1573  & 0.138 & 6.207 & 552 & 1282
           & $ 5.2 \times 10^{-14}$ & 70908 & 7828 & 80309  \\
16 & 2753  & 0.165 & 6.207 & 475 & 1282
           & $ 5.7 \times 10^{-14}$ & 61543 & 6803 & 71099  \\
24 & 3703  & 0.178 & 6.207 & 443 & 1282
           & $ 4.3 \times 10^{-14}$ & 57792 & 6374 & 67869  \\
32 & 4725  & 0.198 & 6.207 & 403 & 1282
           & $ 5.3 \times 10^{-14} $ & 52961 & 5864 & 63550  \\
40 & 6524  & 0.211 & 6.207 & 378 & 1282
           & $ 6.0 \times 10^{-14}$ & 50301 & 5581 & 62406
\end{tabular}
\end{center}
\caption{The execution time for a Pentium 4 (2 GHz) node to compute
  12 columns of quark propagators, versus the number
  of projected low-lying eigenmodes.
  The parameters for the test are :
  the lattice size is $8^3\times 24$; $\beta=5.8$;
  the degree of Zolotarev rational polynomial is $ n=16 $;
  the number of bare quark masses is $ N_m=12 $
  and $ ma \ge 0.06 $;
  each projected eigenmode satisfies
  $ \|(H_w^2-\lambda^2) |x\rangle \|< 10^{-13}$; and
  the stopping criterion for inner and outer CG loops
  is $ \epsilon=10^{-11} $.
}
\label{tab:perf1}
\end{table}

In Table \ref{tab:zolo_n}, we measure the precision of exact chiral
symmetry $ \sigma $ (\ref{eq:sigma}) versus the degree ($n$) of
Zolotarev optimal rational polynomial. The values of $ \sigma $
listed in the second column of Table \ref{tab:zolo_n} are the maxima in
the nested CG loops.
The execution time and the iterations of the nested CG loops
are also listed. Evidently, the precision of exact chiral
symmetry $ \sigma $ is quite different from the stopping criterion
$ \epsilon = 10^{-11} $ for inner and outer CG loops,
since $ \sigma $ can be much bigger or smaller than $ \epsilon $,
as shown in Table \ref{tab:zolo_n}, as well as in
Tables \ref{tab:perf1} and \ref{tab:perf2}.
It is clear that the necessary condition for preserving exact
chiral symmetry to a very high precision is to use a higher
degree ($n$) Zolotarev rational polynomial for $ (H_w^2)^{-1/2} $.
In Ref. \cite{Chiu:2002eh}, tables are provided for looking up
which degree $ n $ is required to attain one's desired accuracy
in preserving the exact chiral symmetry on the lattice,
versus the parameter $ b = \lambda_{max}^2 /\lambda_{min}^2 $
of a given gauge configuration.

\begin{table}
\begin{center}
\begin{tabular}{c|cccccc}
  Zolo.   &  $\chi$ sym. & \multicolumn{2}{c}{CG iters.} &
        \multicolumn{1}{c}{CG}   & $ D^{\dagger} $ mult. & Total \\
 degree   &  $\sigma$(max.) & inner & outer & time & time & time \\
\hline
\hline
 4 & $ 1.7 \times 10^{-4} $ & 367 & 286 & 82393  & 4493 & 86885   \\
 8 & $ 1.5 \times 10^{-8} $ & 402 & 288 & 111247 & 6003 & 117250  \\
10 & $ 2.9 \times 10^{-10}$ & 408 & 288 & 120940 & 6520 & 127460  \\
12 & $ 6.4 \times 10^{-12}$ & 411 & 288 & 129188 & 6962 & 136150  \\
16 & $ 1.4 \times 10^{-13}$ & 414 & 288 & 148654 & 8006 & 156660
\end{tabular}
\end{center}
\caption{
The precision of exact chiral symmetry $ \sigma $
versus the degree ($ n $) of the Zolotarev rational polynomial.
The test is performed at a Pentium 4 (2 GHz) node, with the
parameters: lattice size=$16^3\times 32$; $\beta=6.0$;
the number of bare quark masses is $ N_m=16 $;
the number of projected eigenmodes is $ k=20 $;
each projected eigenmode satisfies
$ \|(H_w^2-\lambda^2) |x\rangle \|< 10^{-13}$;
$ b=\lambda_{max}^2/\lambda_{min}^2=1086$;
and the stopping criterion for the CG loops is $ \epsilon=10^{-11} $.}
\label{tab:zolo_n}
\end{table}

In Table \ref{tab:perf2}, we list the execution time (in unit of second)
of a Pentium 4 (2 GHz) node to compute 12 columns of quark
propagators, versus the size of the lattice.
The parameters for the test are:
the degree of Zolotarev rational polynomial is $ n = 16 $,
the number of bare quark masses is $ N_m = 16 $ and $ ma \ge 0.02 $,
each projected eigenmode satisfies
        $\|(H_w^2-\lambda^2) |x\rangle \|< 10^{-13}$, and
the stopping criterion for the inner and the outer CG loops is
$ \epsilon = 10^{-11} $. From the last entry of the last row, we can
estimate that a Pentium 4 (2 GHz) node takes about 24 days to complete
12 columns of quark propagators (for 16 bare quark masses)
for one gauge configuration at $ \beta = 6.0 $
on the $ 16^3 \times 32 $ lattice.
In other words, if we have 12 nodes and let each one of them work on
one column of $ D^{-1} $, then we can complete the quark
propagators for one gauge configuration in two days.
Since our system consists of 64 nodes,
so we can compute quark propagators
at a rate more than five gauge configurations per two days.

\begin{sidewaystable}
\begin{center}
\begin{tabular}{c|ccccccccccccc}
& & \multicolumn{2}{c}{projections} & & & $\chi$ sym& inner CG & outer CG &
        CG & $ D^{\dagger} $ mult. & disk I/O & Total \\
Lattice & $\beta$ & \# & time & $ \lambda_{min}$ & $\lambda_{max}$ &
        $\sigma$(max.) & ave. iters. & tot. iters.
        & time & time  & time & time \\ \hline\hline
$ 8^3\times 24$ & 5.8 & 32 & 4725 & 0.198 & 6.207
        &  $ 5.4 \times 10^{-14} $ &
        403 & 1322 & 54384 & 7804 & 0  & 66913 \\
$10^3\times 24$ & 5.8 & 30 & 7803 & 0.152 & 6.204 &
        $ 6.4 \times 10^{-14}$ &
        519 & 1943 & 191861 & 18626 & 0 & 218290 \\
$12^3\times 24$ & 5.8 & 30 & 13258 & 0.129 & 6.211 &
        $ 9.8 \times 10^{-14}$ &
        608 & 2840 & 574226 & 38234 & 0 & 625718 \\
$16^3\times 32$ & 6.0 & 20 & 74937 & 0.215 & 6.260 & $ 3.3 \times 10^{-13}$
        &  370 & 3968 & 1866172 & 87890 & 66976 & 2095975
\end{tabular}
\caption{
  The execution time (in unit of second) of a Pentium 4 (2 GHz) node to
  compute 12 columns of quark propagators,
  versus the size of the lattice.
  The parameters for the test are:
  the degree of Zolotarev rational polynomial is $ n=16 $,
  the number of bare quark masses is $ N_m=16 $ and $ ma \ge 0.02 $,
  the precision of each projected eigenmode satisfies
             $\|(H_w^2-\lambda^2) |x\rangle \|< 10^{-13}$, and
  the stopping criterion for inner and outer CG loops is
  $ \epsilon=10^{-11} $.}
\label{tab:perf2}
\end{center}
\end{sidewaystable}


\section{Conclusions}

In this paper, we outline the essential features of a Linux PC
cluster (64 nodes) which has been built at National Taiwan University,
and discuss how to optimize its hardware and software for
lattice QCD with exact chiral symmetry. At present, all nodes 
are working around the clock on lattice QCD computations.

With Zolotarev optimal rational approximation to $ ( H_w^2 )^{-1/2} $,
projections of high and low-lying eigenmodes of $H_w^2$,
the multi-mass CG algorithm, the SSE2 acceleration,
and our simple scheme of memory management,
we are able to compute quark propagators
of $ 16 $ bare quark masses on the $16^3 \times 32$ lattice,
with the precision of quark propagators up to $ 10^{-11} $
and the precision of exact chiral symmetry up to $ 10^{-12} $,
at the rate of 2.5 gauge configuration ($ \beta = 6.0 $) per day,
with our present system of 64 nodes.
This demonstrates that an optimized Linux PC cluster
can be a viable computational system to extract
physical quantities from lattice QCD
with exact chiral symmetry \cite{Chiu:2002xm,Chiu:2002rk,Chiu:2002fy}.

The speed of our system is higher than 70 Gflops, and
the total cost of the hardware is less than US\$60000.
This amounts to price/performance ratio better than \$1.0/Mflops for
64-bit (double precision) computations.
The basic idea of optimization is to let each node work independently
on one of the 12 columns of the quark propagators
(for a set of bare quark masses), and also use
the hard disk as the virtual memory for the vectors
in the outer CG loop, while the CPU is working on the inner CG loop.
Our simple scheme of memory management for the nested CG loops may
also be useful to other systems.

In future, we will add more nodes to our system, and will also work
on larger lattices, say $ 24^3 \times 48 $. Then one Gbyte memory at each
node is not sufficient to accommodate all relevant vectors in the
inner CG loop, even for 12 Zolotarev terms. However,
there are several ways to circumvent this problem.
First, our memory management scheme is quite versatile, which is more
than just for swapping the vectors at the interface of inner and
outer CG loops. In fact, it can handle any number of Zolotarev terms for
any lattice size, and can automatically minimize disk I/O at any step of
the nested CG loops, according to the amount of physical memory of a node.
As long as the percentage of the disk I/O time is less than 30\%, it is
still a better option than distributing the nested CG loops across
the nodes and performing parallel computations (with MPI)
through a fast network switch, since the communication overheads
is expected to be more than 30\% of the total time,
especially for a system of 100 nodes or more.
Secondly, we can increase the amount of memory at each node,
which depends on the specification of the motherboard as well as
the price and the capacity of the memory modules.
Finally, we can also exploit algorithms \cite{Neuberger:1998jk}
which only use five vectors
rather than $ 2 n + 3 $ vectors
for the inner CG loop, or the Lanczos algorithm as described in
Ref. \cite{Borici:1999ws}. Now it is clear that a Linux PC cluster is a
viable platform to tackle lattice QCD with exact chiral symmetry 
even for a large lattice ( e.g., $ 32^3 \times 64 $), though
more studies are needed before one reaches an optimal design
for dynamical quarks.  

\bigskip
\bigskip
\flushpar
{Note added in proof:}

\bigskip
\noindent

Recently, it has been shown \cite{Chiu:2003ub} that the speed of  
Neuberger's double pass algorithm \cite{Neuberger:1998jk}
for computing the matrix-vector 
product $ R^{(n-1,n)} (H_w^2) \cdot Y $ is almost independent of the 
degree $ n $ of the rational polynomial, and it is faster than 
the single pass for $ n > 13 $ (for Pentium 4 with SSE2).  
Thus the single pass has been replaced with the double pass 
algorithm in our Linux PC cluster.

\bigskip
\bigskip
\flushpar
{\bf Acknowledgement}
\bigskip

\noindent

This work was supported in part by the National Science Council,
ROC, under the grant number NSC90-2112-M002-021.

\eject




\end{document}